\documentclass[twocolumn,superscriptaddress,aps,prb]{revtex4}
\usepackage[latin9]{inputenc}
\usepackage{color}
\usepackage{amsmath}
\usepackage{graphicx}
\usepackage{amssymb}

\newcommand{\nio}{Na$_2$IrO$_3$}
\newcommand{\lio}{Li$_2$IrO$_3$}

\newcommand{\rucl}{$\alpha$-RuCl$_3$}

\newcommand{\w}{\omega}
\newcommand{\e}{\epsilon}

\newcommand{\ket}[1]{\left| #1 \right\rangle}
\newcommand{\bra}[1]{\left\langle #1 \right|}

\newcommand{\Ztwo}{\mathbb{Z}_2}
\newcommand{\Zthree}{\mathbb{Z}_3}

\newcommand{\HK}{\mathcal{H}_{\rm K}}
\newcommand{\HM}{\mathcal{H}_{\rm u}}

\newcommand{\HMh}{\mathcal{H}_{\rm \hat{u}}}

\newcommand{\Hz}{\mathcal{H}_{0}}

\newcommand{\rr}{\kappa}  % ration of Kekule couplings

\newcommand{\Kkm}{Kekul\'e-modulated Kitaev model}
\newcommand{\kkm}{kekul\'e-modulated Kitaev model}
\newcommand{\Kek}{Kekul\'e}
\newcommand{\kek}{kekul\'e}

\DeclareMathOperator{\ii}{i}

\begin{document}

\title{
Bound states of fractionalized excitations
in a modulated Kitaev spin liquid
}

\author{Hugo Th\'eveniaut}
\affiliation{Department of Theoretical Physics, KTH-Royal Institute of Technology, Stockholm, SE-10691 Sweden}
\affiliation{Institut f\"ur Theoretische Physik, Technische Universit\"at Dresden,
01062 Dresden, Germany}
\author{Matthias Vojta}
\affiliation{Institut f\"ur Theoretische Physik, Technische Universit\"at Dresden,
01062 Dresden, Germany}

%%%%%%%%%%%%%%%%%%%%%%%%%%%%%%%%%%%%%%%%%%%%%%%%%%%%%%%%%%%%%%%%%%%%%%%

\begin{abstract}
Fractionalization is a hallmark of spin-liquid behavior; it typically leads to response functions consisting of continua instead of sharp modes. However, microscopic processes can enable the formation of short-distance bound states of fractionalized excitations, despite asymptotic deconfinement. Here we study such bound-state formation for the $\Ztwo$ spin liquid realized in Kitaev's honeycomb compass model, supplemented by a {\kek} distortion of the lattice. Bound states between flux pairs and Majorana fermions form in the Majorana band gaps. We calculate the dynamic spin susceptibility and show that bound states lead to sharp modes in the magnetic response of the spin liquid, with the momentum dependence of the corresponding spectral weight encoding internal symmetry of the bound state. As a byproduct, we also show that isolated fluxes may produce Majorana bound states at exactly zero energy. Generalizations and implications of the results are discussed.
\end{abstract}

\date{May 24, 2017}


\maketitle
%%%%%%%%%%%%%%%%%%%%%%%%%%%%%%%%%%%%%%%%%%%%%%%%%%%%%%%%%%%%%%%%%%%%%%%

\section{Introduction}

Strongly frustrated magnets can host a variety of unconventional states of matter.\cite{ramirez94,starykh13} Particularly fascinating are spin liquids and their descendants. Spin liquids are strongly entangled low-temperature states of local moments which do not display magnetic order and do not break any symmetries of the underlying lattice.\cite{balents10}

Typically, spin liquids display the phenomenon of fractionalization: Elementary excitations come with quantum numbers which cannot be constructed from electrons or holes, and local operators can create only multiples (often pairs) of such deconfined excitations, such that locally created excitations decay into fractionalized constituents. As a result, measurable response functions, like the dynamic spin structure factor, display multi-particle continua instead of sharp modes. In fact, the absence of sharp modes is often taken as experimental evidence for spin-liquid behavior.\cite{han12}

However, this reasoning can be spoiled if fractionalized excitations form short-distance bounds states; this is possible despite long-distance deconfinement. Such bound states may arise from attractive interactions between the fractionalized excitations. Hence, it is conceivable to have a deconfined spin liquid whose dominant response arises from bound states and consists of sharp-mode peaks. This highly unusual situation has been studied relatively little in the theory literature,\cite{knolle2,bound_ll} but may be relevant for real materials.
For instance, spin-liquid physics has been proposed to be relevant for underdoped cuprates,\cite{pwa87,ss_pg} but clear-cut evidence for fractionalization is missing, and neutron scattering experiments are typically interpreted in terms of sharp modes.\cite{birgeneau06,jt06,mv09}

It is the purpose of this paper to study the formation of bound states of emergent fractionalized spin-liquid constituents in a solvable setting. To this end, we utilize Kitaev's compass model on the honeycomb lattice which, in its pristine form, realizes a $\Ztwo$ spin liquid.\cite{kitaev06} Its elementary excitations are gapless dispersive ``matter'' Majorana fermions and gapped immobile $\Ztwo$ gauge fluxes (or ``visons'').
By now, a number of honeycomb-lattice magnets have emerged as candidates for realizing dominant Kitaev interactions,\cite{Cha10} most prominently \nio,\cite{Sin10,Liu11,Sin12} different polytypes of \lio,\cite{Sin12,takayama15} and \rucl.\cite{plumb14,sears15} These materials display long-range magnetic order at low temperatures, likely due to the presence of additional Heisenberg interactions,\cite{Cha10,Cha13,kee14,kimchi14,rachel14,perkins14} but it has been suggested that pressure or doping may be used to suppress magnetic order and to induce spin-liquid behavior.

Here we employ the Kitaev model as a playground for bound-state formation.\cite{bound_ll} As well-defined bound states require the existence of energy gaps in the excitation spectrum, we induce dispersion gaps by imposing a \kek-type modulation \cite{chamon00} of coupling constants in the Kitaev model, i.e., we consider a lattice with a periodic $\sqrt{3}\times\sqrt{3}$ superstructure modulation. We show that bound states between matter fermions and isolated fluxes as well as between matter fermions and flux pairs exist in the resulting $\Ztwo$ spin liquid. These bound states are spatially localized (because the fluxes are localized) and come with  different symmetries w.r.t. point-group operations.
We calculate the dynamic spin structure factor and show that bound states involving flux pairs are visible spectroscopically; in fact, the sharp-mode response of the bound states may dominate the spectrum despite deconfinement. The momentum dependence of the sharp-mode spectral weight allows one to deduce the spatial symmetries of the bound states.
We discuss how to detect deconfinement despite low-energy bound-state formation, and we highlight implications for more realistic models and for carrier-doped spin liquids.

We note that bound states between matter fermions and flux excitations were discussed before for the field-induced non-abelian phase of the Kitaev model.\cite{knolle2,lahtinen14}

The body of the paper is organized as follows:
In Section~\ref{sec:model} we introduce the {\kkm} together with its Majorana representation. Section~\ref{sec:noflux} summarizes the excitation spectrum in the flux-free case. Section~\ref{sec:flux} demonstrates the bound-state formation on the level of Majorana-fermion states while Section~\ref{sec:susc} presents the results for bound-state spectroscopy utilizing the dynamical spin susceptibility. Broader implications are discussed in the summary section.
%

%%%%%%%%%%%%%%%%%%%%%%%%%%%%%%%%%%%%%%%%%%%%%%%%%%%%%%%%%%%%%%%%%%%%%%%

\section{\Kkm}
\label{sec:model}

\subsection{Kitaev model}

The Kitaev model\cite{kitaev06} describes spin-1/2 degrees of freedom at sites $i$ of a honeycomb
lattice which interact via Ising-like nearest-neighbor exchange interactions $J^\alpha$. The
anisotropy direction in spin space, $\alpha=x,y,z$, is coupled to the bond direction
in real space, often dubbed ``compass interaction''.\cite{compass_rmp}
Allowing for spatially varying couplings the Hamiltonian reads
\begin{equation}
\label{hk}
\HK =
-\sum_{\langle ij\rangle_x} J_{ij}^x \hat{\sigma}_i^x \hat{\sigma}_j^x
-\sum_{\langle ij\rangle_y} J_{ij}^y \hat{\sigma}_i^y \hat{\sigma}_j^y
-\sum_{\langle ij\rangle_z} J_{ij}^z \hat{\sigma}_i^z \hat{\sigma}_j^z
\end{equation}
where $\hat{\sigma}_j^{\alpha}$ are Pauli matrices, and $\langle ij \rangle_\alpha$ denotes an $\alpha=x,y,z$ bond as in Fig.~\ref{fig:latt}.
In the homogeneous case $J_{ij}^x=J^x$, $J_{ij}^y=J^y$, $J_{ij}^z=J^z$.
For isotropic couplings, $J^x=J^y=J^z\equiv J$, the model then possesses a $\Zthree$ symmetry of
combined real-space and spin rotations.

The Kitaev model is exactly solvable thanks to the existence of an extensive set of conserved quantities, corresponding to $\Ztwo$ fluxes which can be defined for every closed loop on the lattice. For each elementary plaquette of the lattice, the corresponding loop operators can be written as
\begin{equation}
\label{eq:wp}
 \hat{W}_p = \hat{\sigma}_1^x \hat{\sigma}_2^y \hat{\sigma}_3^z \hat{\sigma}_4^x \hat{\sigma}_5^y \hat{\sigma}_6^z
\end{equation}
with eigenvalues $W_p=\pm1$. For periodic boundary conditions there are, in addition, two ``topological'' loop operators $\hat{W}_{1,2}$ that wrap around the torus. Flux conservation implies that the Hilbert space of the Kitaev model can be divided into flux sectors, specified by the set of $\{W_p\}$. The ground state is located in the flux-free sector;\cite{kitaev06,lieb} this remains true in the presence of the {\kek} modulation considered below.

\subsection{{\Kek} modulation}

The solubility of the Kitaev model relies only on threefold lattice coordination, together with the spin structure of the interactions. Hence, variants of the Kitaev model with vacancies,\cite{willans10} random or inhomogeneous couplings,\cite{zschocke15,majoll} or different lattice geometries, both in two\cite{yao07,baskaran09,kiv1,kamfor10} and three\cite{siyu08,mandal09,trebst14} space dimensions, remain solvable and have been discussed.

Here we impose a modulation of the exchange interactions according to a $\sqrt{3}\times\sqrt{3}$ superstructure on the honeycomb lattice, with the goal to induce spectral gaps. This {\kek} modulation, initially discussed in the context of carbon nanotubes,\cite{chamon00} triples the unit cell, but preserves the $\Zthree$ symmetry of combined real-space and spin rotations.\cite{kek_moessner}

\begin{figure}[t]
\includegraphics[width=0.47\textwidth]{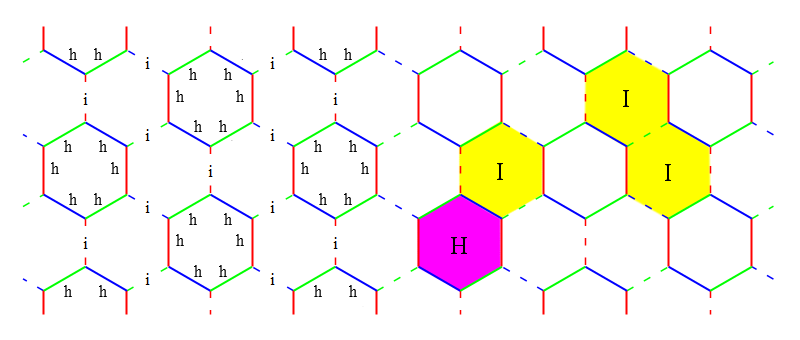}
\caption{
Left: {\Kek} modulation of the Kitaev model, showing h (hexagon, solid) and i (isolated, dashed) bonds according to Eq.~\eqref{eq:bonds}. The $\sqrt{3}\times\sqrt{3}$ superstructure triples the unit cell and generates two types of plaquettes. The colors label the $x/y/z$ bond flavors as in the standard Kitaev model.\cite{kitaev06}
Right: Flux configurations arising from flipping an h (top right) or a i (bottom left) bond.
}
\label{fig:latt}
\end{figure}

The spatial pattern of exchange interactions in the {\kkm} is shown in Fig.~\ref{fig:latt}. The modulation is parameterized by a ratio of coupling constants, $\rr$, such that the exchange couplings obey
\begin{align}
\label{eq:bonds}
J_{ij}^\alpha &= J^\alpha \mbox{ on h bonds} \notag\\
J_{ij}^\alpha &= \rr J^\alpha \mbox{ on i bonds}
\end{align}
Hence, $\rr=0$ corresponds to decoupled hexagons, $\rr=1$ is the original honeycomb lattice, and $\rr=\infty$ represents decoupled dimers. We will be mostly interested in the isotropic case, where we choose $J^x=J^y=J^z\equiv J$, i.e., the couplings on the h bonds, as energy unit.

The enlarged unit cell implies that there are two different types of plaquettes and hence two types of flux excitations; for reference we denote a plaquette formed by six h bonds as H plaquette, while a plaquette with three h and three i bonds is dubbed I plaquette. Flux pairs can be created by flipping either a h bond (creating one H and one I flux) or an i bond (creating two I fluxes), see Fig.~\ref{fig:latt}.

\subsection{Majorana representation}
\label{subsec:Mr}

For explicit calculations we utilize Kitaev's spin representation \cite{kitaev06} with four Majorana fermions per site $\hat{b}^x$, $\hat{b}^y$, $\hat{b}^z$ and $\hat{c}$. Defining $\hat{\sigma}_i^\alpha=i \hat{b}_i^\alpha \hat{c}_i$, the original Hamiltonian in Eq.~\eqref{hk} can be mapped to
\begin{equation}
\label{hm}
\HMh = \ii \sum_{\langle ij\rangle}J^\alpha_{{ij}} \hat{u}_{ij}\hat{c}_i \hat{c}_j,
\end{equation}
where $\hat{u}_{ij}\equiv i\hat{b}_i^{\alpha_{ij}}\hat{b}_j^{\alpha_{ij}}$,
$\hat{u}_{ij}=-\hat{u}_{ji}$ with site $i$ on sublattice $A$, and the summation is over all nearest-neighbor bonds. The $\hat{u}_{ij}$ represent conserved quantities, with
eigenvalues $u_{ij}=\pm 1$, and a given set $\{u_{ij}\}$ reduces the
Hamiltonian to a bilinear in the $\hat{c}$ Majorana operators:
\begin{equation}
\label{hmflux}
\HM = \frac{\ii}{2}\left(\hat{c}^T_A \, \hat{c}^T_B\right) \begin{pmatrix}
0 & M \\
-M^T & 0
\end{pmatrix}
\begin{pmatrix}
\hat{c}_A \\
\hat{c}_B
\end{pmatrix}.
\end{equation}
Here $M$ is an $N \times N$ matrix with elements $M_{ij}=J^\alpha_{ij} u_{ij}$, with $N$ the number of unit cells, and $\hat{c}_{A(B)}$ is a vector of $N$ Majorana operators on the $A(B)$ sublattice. Hence the problem takes the form of non-interacting ``matter'' Majorana fermions coupled to a static $\Ztwo$ gauge field.

For every flux sector, described by a set of set of $\{u_{ij}\}$, the Majorana hopping model $\HM$ can be diagonalized on finite-size lattices using standard techniques, and we refer the reader to the literature.\cite{kitaev06,willans10,zschocke15} The result is a model of free canonical fermions representing the matter degrees of freedom, with
\begin{equation}
\label{hmsvd}
\HM = \sum^N_{m=1} \e_m ( 2 \hat{a}^\dagger_{m}\hat{a}_{m}-1)
\end{equation}
where the mode energies $\e_m$ are non-negative. For the flux-free unmodulated isotropic Kitaev model, momentum $\vec k$ takes the role of the quantum number $m$, and the $\e_{\vec k}$ resemble the positive-energy part of the spectrum of graphene, with Dirac points at momenta $K,K'=\pm(4\pi/(3\sqrt{3}),0)$.

%%%%%%%%%%%%%%%%%%%%%%%%%%%%%%%%%%%%%%%%%%%%%%%%%%%%%%%%%%%%%%%%%%%%%%%

\section{Phase diagram and Majorana spectrum in the flux-free sector}
\label{sec:noflux}

We now turn to the properties of the \kkm. To set the stage, we discuss the spectrum and dispersion of the matter Majorana fermions in the flux-free sector. We allow for coupling anisotropies, and consider the case $J^x=J^y\neq J^z$. In general, there are now three positive-energy bands of matter Majorana fermions due to the tripled unit cell.

\begin{figure}[t]
\includegraphics[width=0.48\textwidth]{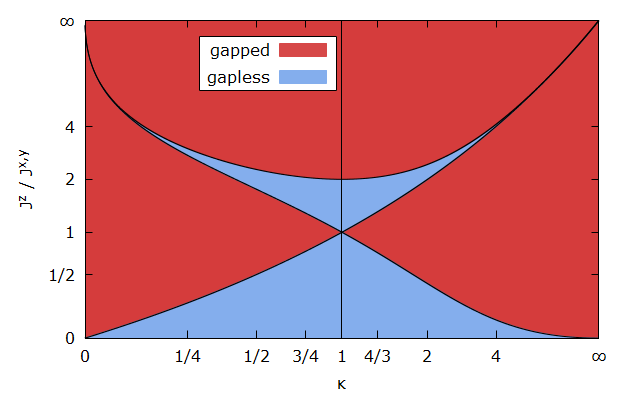}
\caption{
Phase diagram of the {\kkm} as function of modulation $\rr$ and anisotropy ratio $J^{x,y}/J^z$ ($J^x\!=\!J^y$), showing regions of gapless and gapped $\Ztwo$ spin liquids. We note that the horizontal axis is linear in $\rr/(1+\rr)$ while the vertical axis is linear in $J^z/(J^x+J^y+J^z)$.
}
\label{fig:phd}
\end{figure}

The effect of a weak superstructure can be considered perturbatively. In the isotropic case, the {\kek} modulation couples the two Dirac points which are gapped out, hence the spectrum is gapped for any $\rr \neq 1$. In contrast, the anisotropic Kitaev model displays gapless Dirac cones which are shifted in momentum space for\cite{kitaev06} $J^z/J^{x,y}<1$ and $1<J^z/J^{x,y}<2$. In these cases, the {\kek} modulation does not couple the Dirac points, such that extended gapless phases remain for $\rr\neq 1$.
We have computed the phase boundaries, and the resulting phase diagram is shown in Fig.~\ref{fig:phd}. Remarkably, the gapless phases extend all the way to $\rr\to0$ and $\rr\to\infty$.

\begin{figure}[b]
\includegraphics[width=0.49\textwidth]{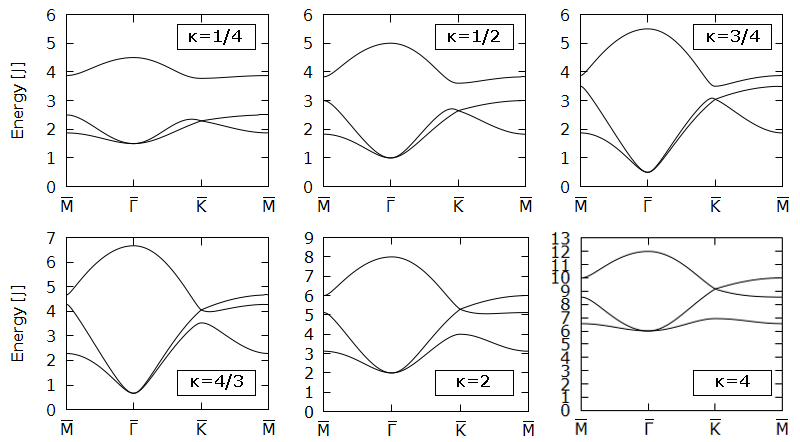}
\caption{
Band structure of the matter Majorana fermion bands, $2\e_{\vec k}$, for isotropic Kitaev couplings and
$\rr=1/4$, $1/2$, $3/4$ (top) and
$\rr=4/3$, $2$, $4$ (bottom).
The momenta refer to the small Brillouin zone of the \kek-modulated lattice.
}
\label{fig:disp}
\end{figure}

\begin{figure}[t]
\includegraphics[width=0.48\textwidth]{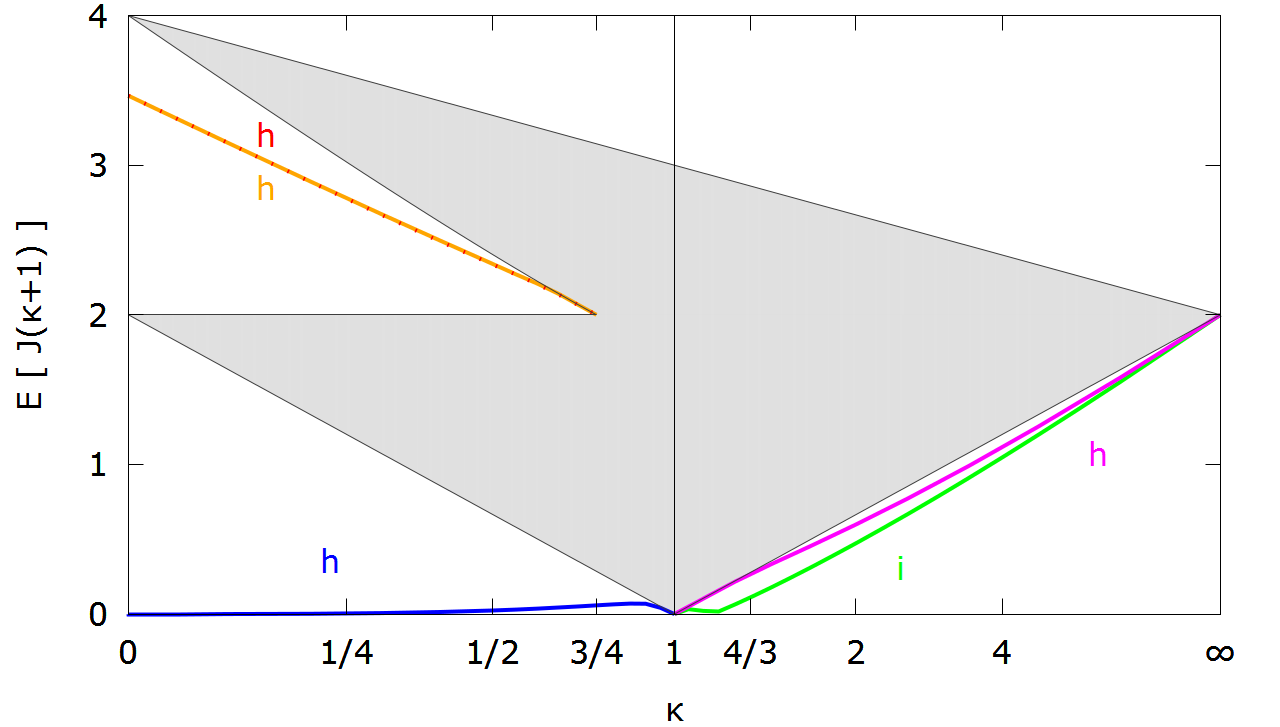}
\caption{
Matter-fermion spectrum of the isotropic {\kkm} as function of $\rr$. The shaded regions indicate the spectrum of the bands in the flux-free case. Its boundaries are given by $4J+2\rr J$, $2J\sqrt{\rr^2-2\rr+4}$, $2J\pm2\rr J$ (left from top to bottom), and $2\rr J - 2J$ (right bottom).
The solid lines represent the energies of the bound states induced by flipping a single h or i bond which creates a flux pair, see Fig.~\ref{fig:latt}.
}
\label{fig:spec}
\end{figure}

In the following we focus on the isotropic case, where we show matter Majorana bandstructures for selected $\rr$ values in Fig.~\ref{fig:disp} and an overview of the spectrum in Fig.~\ref{fig:spec}. As noted, all bands are gapped for $\rr\neq 1$.
For $\rr>3/4$ the three Majorana bands overlap, whereas an inter-band gap opens for $\rr<3/4$.
In the limit of $\rr\to 0$, the spectrum is that of decoupled hexagons, with three discrete local modes, two of which are degenerate, see Fig.~\ref{fig:spec}.
In contrast, the limit $\rr\to\infty$ corresponds to isolated dimers, with a threefold degenerate local mode. Interestingly, at leading order in $1/\rr$ the spectrum corresponds to that of a Kagome lattice.

We note that the transition from $\rr<1$ to $\rr>1$ is accompanied by a band inversion, such that the band structure for $\rr>1$ is characterized by a non-trivial $\Ztwo$ topological index.\cite{kekule_z2} This implies the existence of topologically protected matter Majorana edge states for $\rr>1$.

%%%%%%%%%%%%%%%%%%%%%%%%%%%%%%%%%%%%%%%%%%%%%%%%%%%%%%%%%%%%%%%%%%%%%%%

\section{Bound-state formation}
\label{sec:flux}

Next we turn our attention to the physics in excited flux sectors. Here we demonstrate the existence of spatially localized states of matter Majorana fermions bound to static fluxes which act as ``impurities''. These impurity bound states arise in one of the gaps of the Majorana spectrum.

\subsection{Flux pair}
\label{sec:fluxpair}

We start with flux configurations consisting of a pair of fluxes on adjacent plaquettes; those are relevant for the zero-temperature spin response.\cite{shank,knolle}

We have computed the matter Majorana spectrum by diagonalizing the hopping problem \eqref{hmflux} on finite-size lattices with periodic boundary conditions. Bound-state formation is signalled by the existence of (one or more) isolated mode energies $\e_m$ in energy windows corresponding to the band gaps of the flux-free case. For a single flipped bond, the same information can be obtained by analyzing the poles of a suitable T matrix.

We find matter Majorana bound states for any $\rr\neq 1$, and their energies are summarized in  Fig.~\ref{fig:spec}.
For $\rr<1$ flipping a single h bond generates one bound state close to zero energy; in addition two (almost degenerate) bound states appear for $\rr<3/4$ within the upper band gap. In the limit $\rr\to 0$ these three bound states correspond to the states of an isolated hexagon threaded by a $\pi$ flux, with energies $0$, $2\sqrt{3}J$, $2\sqrt{3}J$. No bound states are obtained by flipping a i bond.
In contrast, for $\rr>1$ bound states are generated by flipping either an h or an i bond; in each case a single bound state emerges below the lower band edge.

\begin{figure}[t]
\includegraphics[width=0.49\textwidth]{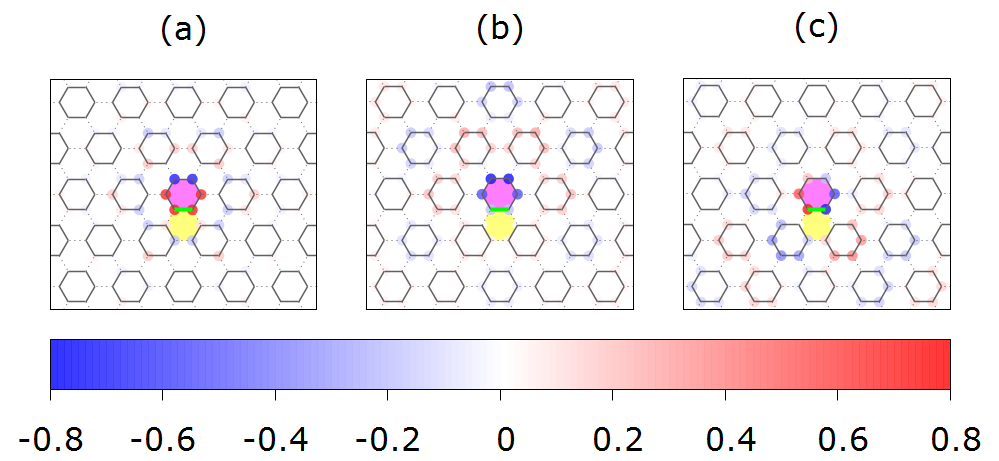}
\caption{
Spatial structure of bound states at $\rr = 0.3$ obtained by flipping a single h bond in the center of the system (shown in green) which creates a pair of fluxes (colored hexagons). The colored circles encode the wavefunction of the localized matter-fermion eigenstate (recall that all eigenstates have real wavefunctions).
The bound-state energies are
(a) $E=0.004J$,
(b) $E=3.472J$,
(c) $E=3.478J$.
}
\label{fig:bs1}
\end{figure}

The wavefunctions of the bound states are portrayed in Fig.~\ref{fig:bs1}. They can be classified according to their signature under reflection at a axis perpendicular to the flipped bond. (We recall that these wavefunctions are not gauge-invariant.)
For $\rr < 1$ the low-energy bound state is even, Fig.~\ref{fig:bs1}(a). The same applies to the lower of the elevated-energy bound states emerging for $\rr < 3/4$, Fig.~\ref{fig:bs1}(b), whereas the higher on is odd, Fig.~\ref{fig:bs1}(c). In all cases, the main weight is located on the H plaquette sharing the flipped bond.
For $\rr > 1$ the bound state obtained from an h flip is even, with the main weight on the I plaquette. Flipping an i bond leaves two mirror symmetries intact (along the bond axis and perpendicular to it), and the bound state is even under the first and odd under the second. Its main weight is equally distributed on the two I plaquettes (not shown).

\begin{figure}[b]
\includegraphics[width=0.45\textwidth]{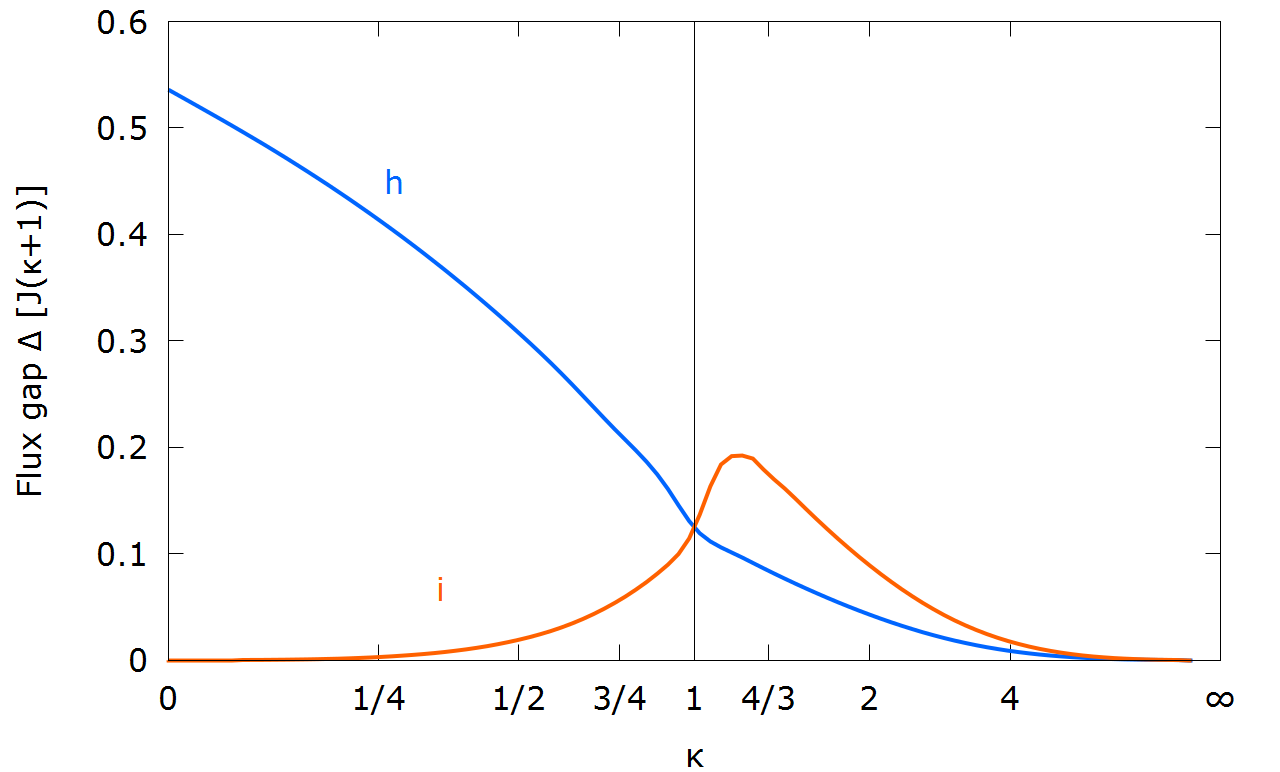}
\caption{
Flux gap $\Delta$ as function of $\rr$: The two lines show the energy cost of flipping a single h (blue) or i (red) bond. The data have been obtained for a system size of $N=900$ where $\Delta=0.25J$; the value in the infinite-system limit is $0.26J$.
}
\label{fig:gap}
\end{figure}

For the computation of the spin structure factor (see below) we will also need the flux gap, i.e., the energy cost of flipping a single bond. The corresponding numerical results are shown in Fig.~\ref{fig:gap}.

We finally note that there are no bound states in the standard Kitaev model, $\rr=1$, even in the anisotropic case $J^x=J^y\neq J^z$ when the spectrum is gapped. However, bound states occur in the presence of the time-reversal-breaking three-site term which is generated by an applied magnetic field.\cite{knolle2,lahtinen14}

\subsection{Single flux}

\begin{figure}[t]
\includegraphics[width=0.48\textwidth]{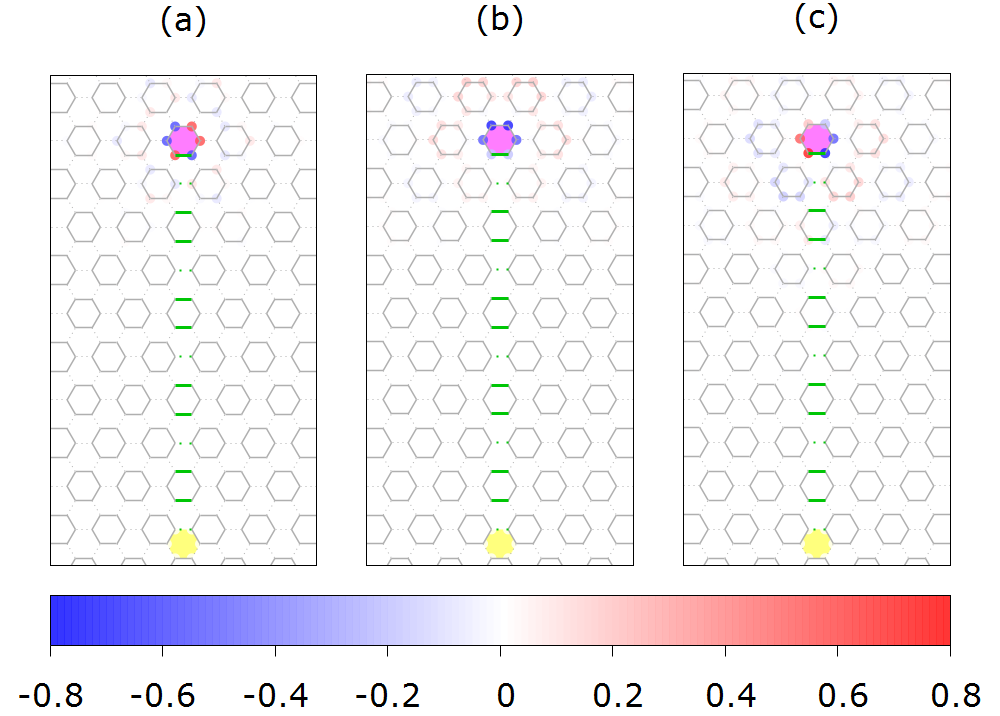}
\caption{
Spatial structure of bound states at $\rr = 0.3$ as in Fig.~\ref{fig:bs1}, but now for a chain of flipped bonds (green) corresponding to two spatially separated fluxes. Remarkably, only fluxes through H plaquettes induce bound states.
The bound-state energies are
(a) $E=0$, (b,c) $E=3.478J$.
}
\label{fig:bs2}
\end{figure}

To complete the physical picture, we now analyze the fermion spectrum in the presence of isolated fluxes. Given that periodic boundary conditions require the total number of plaquette fluxes to be even, i.e., the product of all $W_p$ from Eq.~\eqref{eq:wp} to be $+1$, we study a configuration with two fluxes placed at a maximum distance of $L/2$. Such configurations correspond to flipping a chain of bonds (``Dirac string'') as shown in Fig.~\ref{fig:bs2}.

Interestingly we find for $\rr<1$ that only isolated H fluxes generate bound states. The configurations shown in Fig.~\ref{fig:bs2} have one H and one I flux, and we find a total of three bound states as in the flux-pair case. Their energetics is interesting: For $\rr\to 0$ the energies are again $0$, $2\sqrt{3}J$, $2\sqrt{3}J$. For finite $\rr$ the system preserves the symmetry of the isolated hexagon, and consequently the effect of finite $\rr$ on the bound states is equivalent to a $J$ renormalization. As a result, the two upper bound states are degenerate, and the lower bound state is {\em exactly} at zero energy. Hence, an isolated H flux in the {\kkm} with $\rr<1$ provides a means to generate zero-energy Majorana bound states. Concerning the (gauge-dependent) wavefunctions, we see that lowest bound state one is now odd under reflection w.r.t. the Dirac-string axis. For a configuration with two isolated H fluxes, we find a total of six bound states (not shown).

In contrast, for $\rr>1$ we observe that isolated I fluxes generate one bound state each whereas H fluxes do not induce bound states. Hence, for a configuration with two distant I fluxes we find two bound states (not shown).

%%%%%%%%%%%%%%%%%%%%%%%%%%%%%%%%%%%%%%%%%%%%%%%%%%%%%%%%%%%%%%%%%%%%%%%

\section{Bound-state spectroscopy}
\label{sec:susc}

So far, we have demonstrated the presence of bound states formed by matter Majorana fermions and $\Ztwo$ gauge fluxes or pairs thereof. The latter bound states are directly visible in spectroscopic probes, and we show this by determining the dynamic spin structure factor for the {\kkm}.

\subsection{Dynamical spin correlations}

Dynamical spin correlations in the Kitaev model\cite{shank} have been explicitly calculated in Ref.~\onlinecite{knolle,knolle2}. Consider the zero-temperature spin correlation function
\begin{equation}
S^{\alpha \beta}_{ij}(t)=\bra{0}\hat{\sigma}_i^\alpha(t)\hat{\sigma}_j^\beta(0)\ket{0}
\end{equation}
where $\ket{0}$ is the many-body ground state. The application of a $\hat{\sigma}_i^\alpha$ operator creates a flux pair in the plaquettes which involve the $\alpha$ bond emanating from site $i$. This leads to the dynamical rearrangement of matter fermions in the modified gauge field akin to a quantum quench.
The spin correlator can be expressed purely in terms of matter fermions in the ground-state flux sector, subject to a perturbation $\hat{V}_\alpha = -2iJ^\alpha c_i c_j$.\cite{shank,knolle} For instance, the off-site correlator reads
\begin{align}
\label{sij}
S^{\alpha \beta}_{ij}(\w) = 2\pi F_{ij}^\alpha \sum_\lambda &\bra{M_0}\hat{c}_i\ket{\lambda} \bra{\lambda}\hat{c}_j\ket{M_0} \\ &\times \delta(\w-(E_\lambda-E_0)) \delta_{\alpha \beta} \delta_{\langle ij \rangle_\alpha},
\nonumber
\end{align}
where $\delta_{\langle ij \rangle_\alpha}$ is non-zero only if $i$ and $j$ are nearest neighbors connected by an $\alpha$ bond, i.e., $S_{ij}$ vanishes beyond nearest neighbors. Further, $\ket{M_0}$ is the matter-fermion ground state of $\Hz$ in the flux-free sector.\cite{zschocke_note} $\sum_{\lambda}$ runs over all matter-fermion states in the two-flux sector, i.e., the eigenstates of $\Hz+ \hat{V}_\alpha$. $E_0$ and $E_\lambda$ are the corresponding many-body energies. The prefactor $F_{ij}^\alpha=\{-1,\ii,-\ii\}$ depending on the spin component. Physically, Eq.~\eqref{sij} expresses that a spin-flip excitation decays into a matter Majorana fermion and a flux pair.

\begin{figure*}[t]
\includegraphics[width=0.99\textwidth]{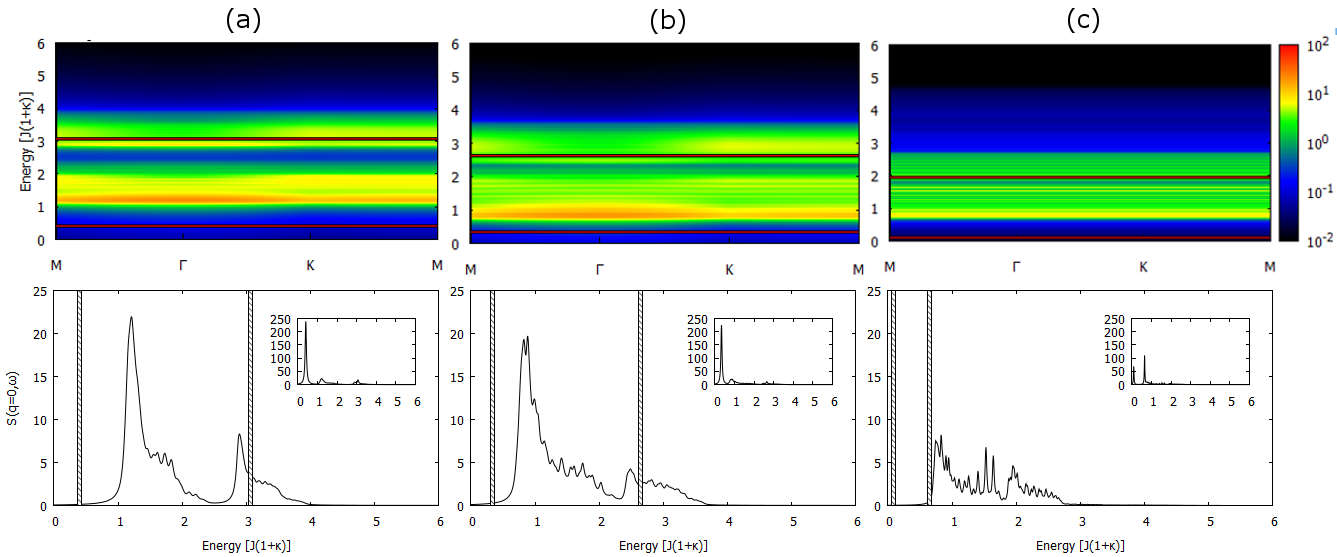}
\caption{
Dynamic spin structure factor for the {\kkm} for coupling ratios (a) $\rr = 0.3$, (b) $\rr=0.5$, (c) $\rr=2$. The upper panels show $S({\vec q},\omega)$ along a path in momentum space using a logarithmic color scale for intensity, with the red lines representing the non-dispersive $\delta$-peak contributions. Lorentzian broadening with $\gamma=0.05J$ has been applied to the rest of the signal.
The lower panels show $S({\vec q}=0,\omega)$, with $\delta$ peaks as vertical lines, while the $\delta$ peaks have been included in the usual broadening scheme in the insets.
}
\label{fig:skw}
\end{figure*}

\subsection{Sources of sharp-mode peaks}

Analyzing Eq.~\eqref{sij} shows that the results crucially depend on whether the ground states in the two flux sectors have (i) the same or (ii) different total fermion parity.\cite{knolle,knolle2} In case (i) non-vanishing contributions to $S^{\alpha \beta}_{ij}(\w)$ arise from excited states $\ket{\lambda}$ with an odd number of matter-fermion excitations while in case (ii) an even number is required.\cite{zschocke_note}
As a result, there are two distinct sources for sharp-mode peaks in the dynamical spin response.\cite{knolle,knolle2}
In case (i), the leading contributions come from excited single-fermion states. Hence, $S(\w)$ reflects the matter-fermion density of states in the two-flux sector, and sharp peaks are caused by energetically isolated single-fermion states -- these are precisely the bound states discussed in Section~\ref{sec:flux} above. The peak energy in $S(\w)$ is then given by the sum of the flux gap and the fermion bound-state energy.
In case (ii), the first contribution is from the zero-fermion state in the two-flux flux sector, and hence always yields a delta peak, located at the flux-gap energy. The next contributions come from two-fermion states, such that the presence of a single fermionic bound state does {\em not} produce a sharp-mode peak.

For the {\kkm} with isotropic couplings, we find for $\rr<1$ that the parity of the two-flux state matches that of the flux-free state both for flipped h and i bonds, i.e., case (i) is realized. In contrast, for $\rr>1$ this only applies to flipped h bonds, while flipping an i bond generates the parity mismatch of case (ii).

\subsection{Results}

We now discuss the dynamic spin structure factor at momentum ${\vec q}$,
\begin{equation}
\label{sqz}
S^{\alpha\alpha}({\vec q},\omega) = \frac{1}{N} \sum_{ij} e^{\ii {\vec q} \cdot (\vec{R}_i-\vec{R}_j)} S^{\alpha \alpha}_{ij}(\omega)
\end{equation}
for the \kkm. We calculate this in a few-mode approximation which has been shown to yield highly accurate results for the standard Kitaev model. The explicit calculation in case (ii) requires to perform a local gauge transformation, and we refer the reader to the literature for details.\cite{knolle2}

Numerical results for $S({\vec q},\omega)$ at zero temperature are shown in Fig.~\ref{fig:skw}. Two striking features are apparent:
First, multiple sharp-mode peaks show up, with some of them even occuring on top of a continuum contribution (despite the fact that bound states are located in a band gap). This can be rationalized as follows:
For $\rr<1$ all $\delta$ peaks in $S({\vec q},\w)$ arise from the bound states discussed in Section~\ref{sec:fluxpair}, with all bound states occurring for flipped h bonds, Fig.~\ref{fig:spec}. Importantly, $S({\vec q},\w)$ involves contributions both from flipped h and flipped i bonds. Since the corresponding flux gaps are different, Fig.~\ref{fig:gap}, the two contributions come with continua which are shifted by a different flux ``offset''. As a result, a bound-state peak from the h-bond contribution may energetically overlap with the continuum from the i-bond contribution -- this is what happens in Fig.~\ref{fig:skw}(a,b).
For $\rr>1$ the spin structure factor contains two sharp-mode peaks of different origin, Fig.~\ref{fig:skw}(c). For a flipped h bond we obtain a bound-state peak, located very close to the lower edge of the continuum, while a flipped i bond leads to the parity mismatch and hence a zero-particle $\delta$ peak at the i flux gap (but no bound-state peak).

Second, there is little momentum dependence in $S({\vec q},\w)$. For the continuum contributions this is similar to the standard Kitaev model where it results from fractionalization and the fact that spin correlators vanish beyond nearest neighbors. In addition, the bound-state peaks are non-dispersive which is a simple consequence of the fluxes (and hence the bound states) being localized.

\begin{figure}[t]
\includegraphics[width=0.48\textwidth]{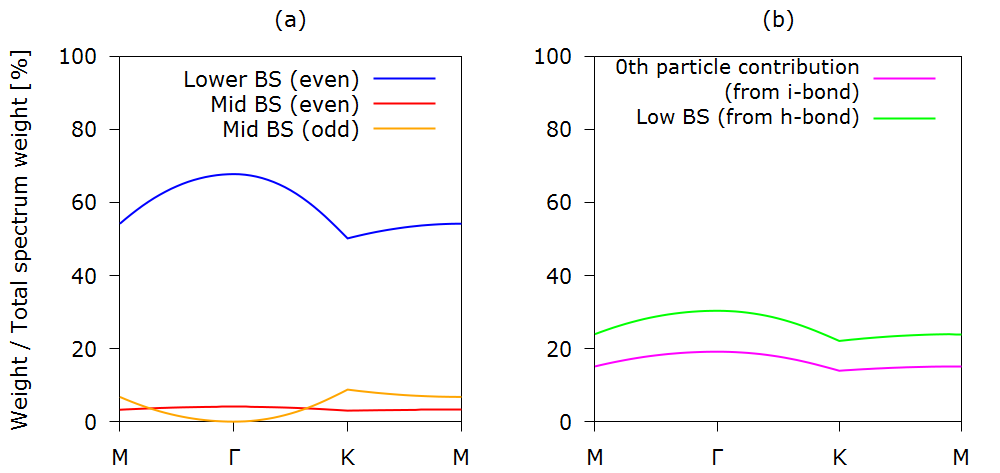}
\caption{
Momentum dependence of relative weight of the $\delta$ peaks in $S({\vec q},\omega)$ at (a) $\rr=0.3$ and (b) $\rr=2.0$. The weights are normalized to the total energy-integrated weight at the respective momentum, $\int d\omega S({\vec q},\omega)$. The line colors match the ones in Fig.~\ref{fig:spec}.
A signature of the odd-symmetry bound state is the vanishing weight at ${\vec q}=\Gamma$ in (a).
}
\label{fig:wgt}
\end{figure}

The weights of the sharp-mode peaks are generically momentum-dependent, as shown in Fig.~\ref{fig:wgt}. First, the weights can be large: For $\rr=0.3$ the peaks account for roughly $2/3$ of the total spectral weight in $S({\vec q},\w)$. Hence, the peaks can dominate the response which also becomes clear from the insets in Fig.~\ref{fig:skw}. Second, the momentum dependence is determined by the internal symmetries of the bound state. This is nicely seen in Fig.~\ref{fig:wgt}(a) where the two bound states which are even under reflection have a finite weight for all momenta, whereas the odd bound state has vanishing weight at $\vec q=\Gamma$.

We can reasonably expect that the response at small non-zero temperatures has the same qualitative properties: The response function will involve a trace over initial states with both fermions and fluxes, but in all cases sizeable matrix elements $\bra{\lambda}\hat{c}_j\ket{M}$ are only obtained if the intermediate state contains, in addition to the excitation effectively created by $\hat{c}_j$, the {\em same} excitations as the initial state $\ket{M}$.\cite{zschocke15}
%(Strictly speaking, corrections to this statement scale with the inverse system size in the limit of small excitations density.)
This implies that all thermal contributions to the low-temperature spin correlator will display the same bound-state peaks, and the main effect of temperature is that of thermal broadening.

%%%%%%%%%%%%%%%%%%%%%%%%%%%%%%%%%%%%%%%%%%%%%%%%%%%%%%%%%%%%%%%%%%%%%%

\section{Discussion and outlook}

In this paper we have studied spin excitations of a {\kkm}. This model provides an explicit and exactly solvable example for a deconfined phase with fractionalized excitation whose dynamic response is nevertheless dominated by sharp-mode peaks which arise via bound-state formation.\cite{bound_ll} We have provided detailed results for the dynamic spin structure factor which we have rationalized via an analysis of spatially localized Majorana fermion states in excited flux sectors.

\subsection{Beyond the Kitaev model}

We now turn to a broader view, and start with physics beyond the solvable Kitaev limit. The gapless spin liquid of the standard Kitaev model has been shown to be stable against small perturbations (like a Heisenberg interaction).\cite{Cha10,Cha13} Two qualitative modifications occur: The $\Ztwo$ flux excitations (visons) become mobile, and spin operator acquires an additional decay channel into two Majorana fermions.\cite{balents16} Upon perturbing the {\kkm} we expect -- by continuity -- that bound states between vison pairs and Majorana fermions continue to exist, but these bound states will themselves become dispersive. Their spatial structure will become more complicated -- in particular, it involves also configurations with the two visons not being on adjacent plaquettes -- but the bound states will continue to contribute sharp modes in the dynamic spin structure factor. Hence, the phenomenon studied in this paper is robust.
In addition, it is also conceivable that bound states between pairs of matter Majorana fermions exist, but this requires a separate analysis which is beyond the scope of this paper.
If the Majorana spectrum does not display spectral gaps, then true bound states can only occur above the upper band edge; alternatively quasi-bound states may exist inside the spectrum. However, neither of the two happen in the standard Kitaev model, $\rr=1$.

For spin liquids with other fractionalization schemes similar bound state formation, e.g., spinon-vison or spinon-spinon, may occur depending on microscopic details, but explicit examples have not been studied to our knowledge. We note that bound-state formation is in fact expected upon approaching a confinement transition: Gauge-field fluctuations produce bound states on all energy scales on the confined side of the transition, and a large confinement length on the deconfined side implies the existence of low-energy bound states. This will studied in future work.

Finally, bound-state formation is relevant beyond insulators. For instance, in doped Mott insulators with deconfinement one may expect charge-neutral spin-$\frac{1}{2}$ spinons and spinless charge-$e$ holons as elementary excitations. Bound-state formation can yield objects with different quantum numbers. In particular, the idea of a fractionalized Fermi liquid (FL$^\ast$) phase\cite{flst} realized in a doped one-band Mott insulator implies the existence of charge-$e$ spin-$\frac{1}{2}$ particles which can be understood as bound states of spinons and holons.\cite{ss_pg}

\subsection{Detecting confinement}

From an experimental point of view, the response from bound states in a deconfined phase is hard to distinguish from the response of a conventional, non-fractionalized phase. In the following we discuss a few aspects of this.
First, a deconfined phase with bound states can be expected to show sharp peaks \textit{and} continua in response functions, see Fig.~\ref{fig:skw}. Notably, such response may also arise from a conventional phase where single-particle peaks and multi-particle continua can coexist. A careful analysis of quantum numbers (e.g. via the response to a magnetic field) may help to distinguish the two cases.
Second, the combination of multiple observables typically provides additional information. For instance, for the Kitaev model it has been shown that Raman scattering mainly probes the spectrum in the ground-state flux sector\cite{knolle_raman} where no bound states occur in the case studied in this paper.
Third, a convincing proof of deconfinement may require to show the existence of emergent gauge-field excitations, for instance via flux trapping in suitable geometries.\cite{flux_trapping}

%%%%%%%%%%%%%%%%%%%%%%%%%%%%%%%%%%%%%%%%%%%%%%%%%%%%%%%%%%%%%%%%%%%%%%%

\acknowledgments

We thank L. Fritz, I. G\"othel, J. Knolle, R. Moessner, S. Rachel, S. Ray, and F. Zschocke
for discussions and collaborations on related work.
This research was supported by the DFG through SFB 1143 and GRK 1621.

%%%%%%%%%%%%%%%%%%%%%%%%%%%%%%%%%%%%%%%%%%%%%%%%%%%%%%%%%%%%%%%%%%%%%%%
%%%%%%%%%%%%%%%%%%%%%%%%%%%%%%%%%%%%%%%%%%%%%%%%%%%%%%%%%%%%%%%%%%%%%%%
%%%%%%%%%%%%%%%%%%%%%%%%%%%%%%%%%%%%%%%%%%%%%%%%%%%%%%%%%%%%%%%%%%%%%%%

%%%%%%%%%%%%%%%%%%%%%%%%%%%%%%%%%%%%%%%%%%%%%%%%%%%%%%%%%%%%%%%%%%%%%%%


\begin{thebibliography}{99}
\bibitem{bound_ll}
The formation of probe-induced bound states in modifications of the Kitaev model has been noticed in Ref.~\onlinecite{majoll} for the special case of triaxial strain.
\bibitem{zschocke_note}
Note that we ignore the fermion parity condition for the ground state\cite{loss,zschocke15} because we are interested in the thermodynamic limit where its effects vanish.
\end{thebibliography}
\end{document}